\documentclass[12pt]{article}
\usepackage{amsmath}
\usepackage{amssymb}
\usepackage{epsfig}
\usepackage{cite}
\usepackage{color,colordvi}

\begin{document}
\vspace{0.00cm}
\begin{center}
{\Large\bf Ultra-High Energy Probes  of  Classicalization} 

\end{center}
%Emergent closed and open string geometry from black holes and species\\
%{\it or shorter}\\
%Emergent closed and open strings from black holes and species\\
%{\it or }\\
%Emergent open strings  and Grand Unification from black holes and species
%} 

%\date{}

%\maketitle

%\begin{center}
%\emph{$^{1        }$ Max-Planck-Institut f\"ur Physik, F\"ohringer Ring 6, \\
  %80805 M\"unchen, Germany } \\

%\vspace{-1.4cm}

\vspace{0.1cm}
%\end{center}

\begin{center}

{\bf Gia Dvali}$^{a,b,c,d}$\footnote{georgi.dvali@cern.ch} and {\bf Cesar Gomez}$^{a,e}$\footnote{cesar.gomez@uam.es}

\vspace{.6truecm}

{\em $^a$Arnold Sommerfeld Center for Theoretical Physics\\
Department f\"ur Physik, Ludwig-Maximilians-Universit\"at M\"unchen\\
Theresienstr.~37, 80333 M\"unchen, Germany}

%\vspace{.2truecm}

{\em $^b$Max-Planck-Institut f\"ur Physik\\
F\"ohringer Ring 6, 80805 M\"unchen, Germany}

{\em $^c$CERN,
Theory Department\\
1211 Geneva 23, Switzerland}

%\vspace{.2truecm}

{\em $^d$Center for Cosmology and Particle Physics\\
Department of Physics, New York University\\
4 Washington Place, New York, NY 10003, USA}

{\em $^e$
Instituto de F\'{\i}sica Te\'orica UAM-CSIC, C-XVI \\
Universidad Aut\'onoma de Madrid,
Cantoblanco, 28049 Madrid, Spain}\\

\end{center}

%\vspace{0.5cm}

\begin{abstract}
\noindent  
 
{\small 
Classicalizing theories  are characterized by  a rapid growth of the scattering cross section.
This growth converts these sort of theories in interesting probes for ultra-high energy experiments even at relatively low luminosity, 
such as cosmic rays or Plasma Wakefield accelerators.  
  The microscopic reason behind   this growth
  is the 
   production of  $N$-particle states, classicalons, 
 that represent self-sustained lumps of soft Bosons. 
   For spin-2 theories this is the quantum portrait of what in the classical limit are known as black holes.   We emphasize  the importance of this quantum picture  which 
  liberates us from the artifacts of the classical geometric  limit  and allows  to  
scan a  much wider landscape of  experimentally-interesting quantum theories.   We identify  a phenomenologically-viable class of  spin-2 theories for which   the growth of 
 classicalon production cross section can be  as efficient  as  to compete with QCD cross section 
 already at  $100$TeV energy, signaling  production of quantum black holes with graviton occupation number  $N \sim 10^4$.  }

\end{abstract}

\thispagestyle{empty}
%\clearpage

%\tableofcontents

\section{Classicalization Versus Wilsonian Physics}

   One of the fundamental goals of high-energy physics is to understand the nature of UV-completion of the Standard Model. 
 In the standard (Wilsonian) paradigm 
 of UV-completion, the new high-energy physics comes in form of weakly-coupled  quantum particles 
that become relevant degrees of freedom at scales shorter than the 
weak interaction length, approximately $10^{-16}-10^{-17}$ cm.   For example, a  low scale supersymmetry 
is a typical representative of such a Wilsonian UV-completion. 
  As it is well-known, the above energy frontier is currently being probed by the LHC experiments.   

 It is natural to ask, what are the prospects of experimentally testing ideas about UV-completion 
 in the collisions at even higher energes?  One rare opportunity is provided by ultra-high energy cosmic rays
 (see \cite{Rev} for a review). Interestingly,  there are also proposals of earth-based accelerators of ultra-high energies, such as  Plasma Wakefield accelerators (see \cite{Allen}). 

 In both of these cases the potential difficulty is a relatively low luminosity, which makes the test 
 of Wilsonian UV-completions exceedingly difficult.    For example,  the high energy cosmic ray 
 collisions in the atmosphere are dominated by soft exchanges and  the statistics for the events with high momentum-transfer is low.  The purpose of the present paper is to identify a class 
of theories, based on the concept of non-Wilsonian self-completion \cite{gia-cesar} by classicalization \cite{class},   for which  the ultra-high energy experiments can represent efficient probes even at relatively low luminosity, due to an efficiently growing cross-section at high energies.   A  potential importance of this general property for compensating the expected  low luminosity of Plasma Wakefield accelerators was noted in \cite{Allen}.

   In order to probe the short-distance physics, one needs events with high momentum-transfer. 
  In weakly-coupled (Wilsonian)   UV-completions of the Standard Model such events are predicted 
  to be extremely rare  as compared to low momentum-transfer soft events mediated by the strong 
  interaction force.  
  The reason is that in any  weakly-coupled UV-completion of the standard model 
  the (all-inclusive)  cross section diminishes with the increase of the center of mass energy $E$. 
  As a result, the cross section is dominated by soft QCD events. 
  This is why, at  least  for the current statistics,  the high energy cosmic rays  do not represent 
 good probes of Wilsonian UV-completions of the standard model. 
 
   Recently, a concept of non-Wilsonian self-UV-completion was introduced  \cite{gia-cesar, class}, 
 in which no new weakly-coupled physics is required  above certain cutoff energy scale  
 $M_*$ with the corresponding cutoff length  $L_* \equiv \hbar /M_*$.  
 In these theories low energy degrees of freedom (e.g., gravitons) naively become strongly interacting 
at distances shorter than $L_*$  where perturbative expansion in $E\, L_*$ breaks down. 
However, in contrast to the Wilsonian picture, this breakdown does not imply the need  for  any 
new weakly-coupled physics that must be integrated-in at distances less than $L_*$. 
 Instead,  the theory cures itself  in the following way. 
 The would-be strongly-coupled particles get replaced by collective weakly-coupled degrees of freedom  with an effective interaction strength suppressed by powers of $1/(L_*E)$. These collective degrees of freedom represent many-particle states of large wavelength. By large wave-length
 we mean the wavelengths that exceed $L_*$.   
   This characteristic  length  is set by an energy-dependent  scale which we denote by $r_*(E)$.   The necessary property is the increase of $r_*(E)$  with $E$, so that 
 $r_*(E)   \, \gg \, L_*$ for  $E \, \gg M_*$.   As a result, the UV-theory  when described in terms of collective  degrees of freedom is weaker and weaker coupled and probes larger and larger distances with 
 increasing $E$. 
  This phenomenon  describes the essence of what was termed as 
    the non-Wilsonian self-completion by  {\it classicalization} \cite{class}.
   Various aspects of this idea where studied in \cite{giadavid} - \cite{sigmamodel}.    
  Intrinsic feature of the classicalization phenomenon, which  makes this picture  phenomenologically distinct from Wilsonian completion  is  the efficient growth of the cross section  at trans-cutoff energies 
  as some positive power  of  $ EL_*$, 
  \begin{equation}
    \sigma \, \propto \, (EL_*)^{\alpha}  \,,  ~~~\, \alpha \, > \, 0
   \label{growth}
   \end{equation}  
 
 % These weakly coupled fields can be viewed as excitations of multi-particle bound-states.  

    This growth can be understood as the result of  creation of  states composed of many soft 
 quanta,  which behave 
 more and more classically at high-energies.  These are so-called {\it classicalons}.  
%  The necessary requirement for such UV-completion 
 %is the existence of a bosonic field (self)sourced by  energy. 
 %The high-energy scattering is then dominated by configurations of this field that include 
 %many soft quanta. Such configurations are generically referred to as {\it classicalons}.  
  Let us briefly describe their essence. 
  %What are the classicalons? 
  In the classical limit ($\hbar \, = \, 0$) classicalons represent static (usually singular) solutions of the classical equations of motion of characteristic radius $r_*(E)$ and energy (mass) $E$.   These parameters appear as integration constants that can take arbitrary values. 
  A well-known example of such solutions is the celebrated Schwarzschild black hole 
  in classical general relativity. 
   However, this geometric picture is only valid in an idealized classical limit.  In reality nature 
   is quantum and $\hbar$ is non-zero.  A quantum theory of black holes and other classicalons 
   was developed in \cite{gia-cesar-alex, number}.  
   According to the latter theory,  these objects  represent self-sustained  bound-states of many bosons of characteristic wave-length $\lambda \, = \, r_*(E)$ and occupation
 number, 
 \begin{equation}
   N(E) \, = \, E \, r_*(E)\, / \hbar \, .
   \label{N}
   \end{equation}
 Due to their large wave-length and derivative coupling,  these bosons interact extremely weakly,  with the effective coupling constant 
 \begin{equation}
 \alpha_{eff}  \, = \, 1/N(E)  \, .
 \label{coupling}
 \end{equation}
 Thus \cite{number},  physics of  classicalons in general,  and black holes in particular, 
 is  a weakly-coupled large-$N$ physics in t'Hooft's sense \cite{tHooft}.  
  This property emerges as the result of maximal packing.  The classicalons  represent maximally packed states 
  per given wave-length.   The  maximized occupation number density results into the oversimplification of the system and effectively converts it into a system with a single characteristics,  
  $N$.  
 In this way, classicalization replaces a would-be strongly coupled physics of few hard quanta  at energy 
 $ E \, \gg \, M_*$ by an extremely weakly-coupled physics in which the same energy is distributed among many soft quanta  of wavelength $r_* \, \gg \, L_*$.  
 
  This picture defines the quantum $N$-portrait of black holes and other classicalons. 
  The  reason for the efficient  production rate of these objects  in high energy particle collisions is the exponential degeneracy of micro-states that over-compensates the usual 
exponential suppression of many-particle states.   At high energies the cross-section 
grows as 
\begin{equation}
\sigma\, \sim \,  (\hbar \, N(E) /E)^2 \, .
 \label{crosssection}
 \end{equation}
 This cross section for large $N(E)$ can be interpreted  as a geometric cross section 
 \begin{equation}
 \sigma \,  \sim \, r_*(E)^2 \, .
 \label{geometric} 
 \end{equation} 

 The relations (\ref{N}), (\ref{coupling}),  (\ref{crosssection}) and (\ref{geometric}) describe the essence of classicalizing theories.   Unlike in ordinary Wilsonian case,  the high energy behavior of these theories , instead of probing short distances, in reality probes large distance physics, 
 due to the fact that the high energy scattering is dominated by production of states 
 with large occupation number of very soft quanta.

 Thus,  deep-UV quantum behavior of  classicalizing theories can be understood in terms of the classical IR dynamics of the same theory!
   For example, the behavior of deep-UV cross section can be derived by 
 finding out the $E$-dependence of the $r_*(E)$ radius of a static source of mass $E$.   
 The radius $r_*$ can be defined as the shortest distance 
 for which the linearized approximation is valid.  This property simplifies enormously the 
 predictive power of the theory for high-energies, since the dependence of  
 $\sigma$ on center of mass energy $E$ can be read-off   by solving the linearized classical equations of motion for a source of the same energy $E$.  

 However, we need to be extremely careful not to be mislead 
 by this simplification.   In order to understand properly the classicalon dynamics 
 we must continuously monitor the information obtained in an idealized classical limit  ($\hbar = 0$) by translating it into the language of the  underlying quantum portrait. 
  Without this guideline, the (semi) classical picture alone can  lead us to wrong conclusions. 
 This becomes obvious, ones we identify the correct classical and semi-classical limits. 
   These  limits  correspond to  taking  
   \begin{equation}
   E \, \rightarrow \, \infty \, , ~~~ \,  L_* \, \rightarrow \, 0\,  ~~~ \, r_* \, = \, {\rm fixed} \, . 
   \label{limit}
   \end{equation} 
  In addition, we may  take $\hbar \, \neq \, 0$ or  $\hbar  \, = \, 0$ depending whether 
  we want to be in semi-classical or classical treatment.  For example, most of (if not all)  the previous semi-classical analysis of the black hole physics  is performed in this limit. 
  
   The quantum $N$ portrait shows that none of these limits are correct approximations. 
This becomes very clear by realizing that  in both limits, irrespective whether 
   we keep  $\hbar$ finite or zero,  the occupation number of quanta becomes infinite, 
   $N \, \rightarrow \, \infty$.  This immediately tells us  that all the subtleties  of  $1/N$ expansion 
   become hidden.   The typical example of the invalidity of this approximation is the application of the semi-classical limit 
   for micro-black holes that can be produced at LHC.  It is obvious that in reality 
   these black holes correspond to the quantum states with $N \,  \sim \, 1$. Thus, to apply to their 
   properties the semi-classical limit ($N \, = \, \infty$) gives invalid  predictions, such as thermality  and democracy of their decay products.  In reality,  the micro-black hole if accessible at LHC  will behave  simply as unstable quantum particles.

   Likewise, for classicalons  (including black holes)  with large $N$ -- that as we shall argue 
 can be observed  in high energy cosmic ray experiments --   the classical limit can serve as an useful guideline.  But at the same time the quantum $N$ portrait makes an essential difference at all the scales, since it allows us to bypass  huge technical complications of extracting information from the classical analysis.  
 
  For example, in the classical limit the question  of black hole formation  is  complicated  with all 
  possible technical subtleties, such as the question of horizon formation. 
  Since in such a limit the quantum nature is  completely hidden, the only remaining  
 characteristics are geometric and one has to be extremely precise  to prove the 
 black hole formation. 
  These technicalities, if the only guideline used is the geometric picture,  make the outcome of high-energy scattering inconclusive.
    
    Instead,  the large $N$ portrait sheds a very powerful light on this process. 
    It makes clear that all the geometric issues, such as  horizon-formation are secondary if we are interested in the question of  formation of large-$N$ quantum states.  For large $N$, this question can be answered  reliably in $1/N$ expansion, which is an excellent approximation for 
    the cases of interest.  
     
     With the large-$N$ quantum portrait in mind,  the questions  whether the  large-$N$ states are formed can be reliably answered already by linearized classical  analysis properly translated into the underlying quantum   language. 
 
   Equipped with the above knowledge,  in the present note we shall focus on a class of theoretically-motivated classicalizing    theories that can be of interest for  high energy cosmic rays  as well as for other high-energy 
but low luminosity experimental searches, with the signatures that can be cross 
correlated with the signatures at LHC. 

 The class of theories of our interest represents UV-modification of Einstein's gravity that
 classicalize  at energies above   $M_* \, \sim $ TeV and thus are theoretically  motivated  
 for solving the hierarchy problem.  They can be viewed as a generalization of 
 large extra dimensional scenario \cite{ADD}, but  we shall abandon the geometrical framework allowing ourselves  to characterize the dynamics in terms of degrees of freedom and their occupation number. 
  As we shall see, this approach since it liberates us from the frame of geometric thinking,   
  allows us to uncover a much wider class of phenomenologically-interesting  possibilities. 
  
   We shall show that in the most efficiently classicalizing scenario, the high-energy 
 growth of the cross section can be as steep as 
 \begin{equation}
\sigma (E)  \, \simeq \,  \pi \, ( E^2  L_*^4 \, ) \, .
\label{sigma}
\end{equation}   
This approximately-quadratic growth takes place until the energy  $E_c \, \equiv \,     (L_*^2m)^{-1}$, where,  $m$ is the mass of the new spin-2 degrees of freedom. 
 Above $E_c$  the  power-law growth of the cross section changes to 
a logarithmic growth according to 
 \begin{equation}
\sigma (E)  \, = \,  \pi \, m^{-2} \, ln^2 \left ({EL_*^2 \over \sqrt{\sigma (E) /\pi} }\right) \,.
\label{sigma}
\end{equation}   
%where,  $m$ is the mass of the new spin-2 degrees of freedom. 
%Above $L_*^{-1}$ and until the energy  $E_c \, \equiv \,     (L_*^2m)^{-1}$ this cross section grows approximately quadratically 
%with $E$, 
%\begin{equation}
%\sigma (E)  \, \simeq \,  \pi \, ( E^2  L_*^4 \, ) \, .
%\label{sigma}
%\end{equation}   
%\begin{equation}
%\sigma (E)  \, \simeq \,  \pi \, ( E^2  L_*^4 \, )\, ln^2 \left ({1 \over  1\, - \, mEL_*^2}\right) \,, 
%\label{sigma}
%\end{equation}   
%Above $E_c$  the  power-law growth of the cross section changes to 
%a logarithmic growth.  
%until the ultra-Planck scale energies  ($\sim \, M_P^2/m$) that are way beyond the current experimental reach. 
We estimate the phenomenological lower bound  on the scale $m$, and 
show  that  for $M_* \, \sim \, TeV$  it  can be as low as $\sim \, 20$MeV. 
This lower bound comes from the combined constraints of supernova cooling and the 
big bang nucleosynthesis.

%temperature $T_{SN} \, = \, 30 $MeV of supernova 1987 A.  The precise bound depends on the number of new graviton species.  
%If the number is large the bound is approximately  $m \, \gtrsim \, 30 T_{SN}$, whereas 
%for smaller number can be relaxed even below $T_{SN}$.    
 
  To conclude,  we show that in the interval of phenomenologically interesting energies, classicalons  with occupation number  as large as $N \sim 10^4$ can be produced in high energy cosmic ray collisions with atmosphere at energies as low as $100$ TeV, with 
  their   production cross-section competing or even exceeding  
the typical low-momentum-transfer QCD cross-section
$\sigma_{QCD}\, \sim 100 mb \, \sim \,  250 $GeV$^{-2}$. 
% The question for which energies this bound gets saturated depends on the 
%value of $M_*$.  For example for $M_* \sim $ TeV,   $\sigma_{BH}$ starts exceeding  $\sigma_{QCD}$ for 
%approximately  $E \sim 100$TeV.   

%For the phenomenologically acceptable lower bounds on $m$ and $M_*$, the above 
%cross section can dominate over the QCD one by factor of $30$ for high-energy 
%cosmic rays in the 100TeV range.  

   Of course,  the possibility of micro black hole production in above-TeV-energy collisions
   is not new and was already suggested in \cite{AADD} and in a number of subsequent papers
   (see, e.g.,  \cite{micro}). 
   This suggestion was based on combining the  intuition about the classical black hole formation in untra-Planckian energies pioneered in earlier papers \cite{tH, ACV, Gross}  with the new input of possibility of lowering the  fundamental Planck mass down to TeV energies \cite{ADD}.
     Therefore,  most of the previous studies were done within the classical geometric frameworks 
   such as  \cite{ADD}, which cannot capture the underlying quantum  large-$N$   
  structure of black holes.  By abandoning the pure geometric limit we manage to identify 
  classes of theories with more efficient growth of the cross section.  The purpose of the present paper 
  is not to confront these models with any particular experimental data, but to point out their
 phenomenological viability and their potential experimental importance for ultra-high energy 
 experiments, both cosmic ray as well as the earth based accelerators.   
   %   Due to efficient growth of the cross section,  the classicalizing theories discussed in this paper represent potential  interest also for
      
      A very interesting representative of such an earth-based ultra-high-energy  accelerator 
is a currently-planned proton-driven  wakefield accelerator \cite{Allen}.  
    These experiments have a potential  of going significantly above the current accelerator  
 capacities, but with relatively low luminosity.  However, for probing classicalizing theories the low luminosity is  
 not a fatal problem since the large cross section can make up for it.    Hence, such ultra-high energy experiments represent potentially very interesting probes for the classicalization phenomenon.

 % In a two-particle scattering of a center of mass energy $\sqrt{s}$,  the cross-section increases 
 % as  $\sigma \, = \, N L_P^2$ and for large $N$  imitates a geometric cross section
 % $\sigma \, \sim \, r_g^2$   of production of a classical object, black hole (or a classicalon  in more general terms) of Schwarzschild radius $r_g$.   This is the picture in 
 % pure Einstein gravity in which the only perturbative messenger of gravitational interaction
  %is a single Einsteinian graviton, and correspondingly the fundamental length is $L_P$. 
 
 % Unfortunately,  such high-energies are not achievable in particle physics colliders, and 
  %for ordinary Einstein gravity  direct collider tests  test for our ideas 

  \section{Spin-2 Classicalizers} 
  
    We  shall discuss the phenomenological implications of classicalizing 
  theories in which the classicalizer fields are spin-2 bosons.    Our aim is to identify 
  the most efficient low-energy classicalizers that could be of potential interest for 
  cosmic ray physics.   These theories are  in the same  universality class as 
 the original large extra dimensions in which fundamental scale of gravity is  $M_* \, \sim $ TeV \cite{ADD}.   In the latter class of theories, above the energy scale set by the inverse 
 compactification radius, the theory is effectively described by  Einstein gravity in $4 + d$ dimensions, with the fundamental Planck length being $L_*$.    Such a theory 
 classicalizes above the energy scale $M_*$.     According to the large-$N$  quantum portrait,  the classicalons of this theory 
 are many-graviton states of wavelength, 
 \begin{equation}
 \lambda \,  =\, r_*(E) \, = L_* \, (E\, L_*)^{{1\over 1+d}} \, , 
 \label{waved}
 \end{equation}
and occupation number  
\begin{equation}
 N \,  =\, \hbar^{-1} E \, r_*(E) \, = \hbar^{-1} (E\, L_*)^{{2+d\over 1+d}} \, .  
 \label{Nd}
 \end{equation}
 Putting it shortly, black holes are graviton Bose-condensates. 
 Given the wave-length, the quantum interaction strength between graviton pairs is given by 
 \begin{equation}
 \alpha_{gr} \, = \,  \hbar \, (E \, L_*)^{-{2 + d \over 1+d}} \, .
  \label{wave}
  \end{equation}
  Notice that all these quantities can be expressed in terms of $N$ as, 
  \begin{equation}
  \lambda \, = \, N^{{1 \over 2+d}} \, L_* \, ~~~\,  \alpha_{gr} \, = \, 1/N \, . 
  \label{DN}
  \end{equation}   
In the limit $\hbar \, =\, 0, ~ L_* \rightarrow 0,  N  \, \rightarrow \, \infty$,  these objects can be viewed as  $4+d$-dimensional Schwarzschild black holes of radius  $r_*$.   
 Thus, physics of black holes is 
 the same large-$N$ physics  that was used by  t'Hooft for $SU(N)$ gauge theories, except the  
role of $N$ is played by graviton occupation number.

  However, the large-$N$ quantum portrait of black holes makes it clear that 
  overemphasizing the geometric meaning is not always very useful. 
First , for  moderate values of $N$ the geometric picture is simply not a good description. 
 Secondly, it illustrates that restricting the analysis to the  configurations that have well-known geometric  interpretation in the $\hbar \, = \, 0$ limit, leaves aside a huge class of potentially interesting cases 
which can be easier to access in high energy experiments and could  have more spectacular experimental signatures than the "canonical" cases which have been studied 
in the geometric  classical limit. 
 
  Therefore, the correct attitude in this search would be to adopt the fully quantum mechanical view 
in which the main role is played by the degrees of freedom and their occupation number,  
with a  geometric interpretation being secondary.

       Gravity  viewed as a quantum field theory  is a theory of a dynamical metric, $g_{\mu\nu}(x)$, which 
   can be represented as a background metric $\langle {g}_{\mu\nu} \rangle$ 
   plus a small perturbation, $\delta g_{\mu\nu}$, 
   \begin{equation}
    g_{\mu\nu} \, = \, \langle {g}_{\mu\nu} \rangle \, + \, \delta g_{\mu\nu} \, .
    \label{metric}
   \end{equation}
% Obviously,  $g_{\mu\nu}$ is a frame-dependent quantity.  In studying gravitational properties of a given localized source for us would be useful to go into the reference frame in which the center of mass of the source is static.    
     We shall be interested in dynamics of the sources on asymptotically-Minkowski spaces, which is an excellent  approximation for  the problems we shall study. 
      
   In such a case, the metric created by a static source  sufficiently  far from it, is
  approximately  Minkowskian  and can be described as, 
   \begin{equation}
   g_{\mu\nu} = \eta_{\mu\nu} \, + \, \delta g_{\mu\nu} \, ,
    \label{almostflat}
    \end{equation}
  where  $\delta g_{\mu\nu} \, \ll \, 1$. 
   The key approach of the quantum $N$-portrait is to think  of such a geometry as a quantum 
   state  (a sort of Bose-condensate)  that is characterized by the  occupation number  $N$,  
   the wavelengths of gravitons $\lambda$ and by their interactions strength $\alpha_{gr} $. 
    Classicalons are then the special states in which all these parameters  assume values that  are fully determined by  $N$,   in such a way that a self-sustainability  condition is satisfied. 
   For example,  the quantum counterparts of Schwarzschild black holes  are given by (\ref{DN}).     
  %  for Einstein gravity in four-dimensions this is achieved when, 
   %\begin{equation}
  % \alpha_{gr} \, = \, 1/N \,,  ~~ \lambda \, = \, \sqrt{N} L_P \, .
  % \label{4N}
  % \end{equation}
  %In this case the quanta form a leaky Bose-condensate.   
   
    The dictionary between the quantum and classical 
   characteristics can be established as follows. 
  
  Metric perturbation   $\delta g_{\mu\nu}$ can be written in terms of  linear perturbations of  canonically-normalized fields that can be classified by their transformation properties with respect 
  to the Poincare group.   Approaching the source from infinity, 
  we shall measure  $\delta g_{\mu\nu}$ which can be found by solving the appropriate 
  linear equations, in which  it is sourced by an effective energy-momentum tensor 
  $T_{\mu\nu}$ of energy $E$.     Assume that this approximation breaks down at some distance $r \, \equiv \, r_*$ for which $\delta g_{\mu\nu}$ becomes order one.  We shall call the corresponding 
  radius, $r_*$ the  {\it classicalization}  scale. 
  
    The quantum mechanical meaning of $r_*$ is that we are dealing with a quantum-mechanical 
    state with graviton occupation number given by  (\ref{N}) and the characteristic wavelength, 
    \begin{equation}
     \lambda \, = \, r_*
   \label{wave}
  \end{equation}
    
  The term  {\it classicalization  radius} has an obvious meaning. It is enough to notice that 
  $N \, \gg  \, 1$  whenever $r_*$ exceeds the Compton wavelength of the source  ($ \hbar /E$). 
  That is, we are dealing with a large occupation number state that can approximately be treated as 
  a classical object.  In a quantum Universe like ours, this is the only consistent  meaning of classicality. 
       
     Also, $r_*$ acquires a classical geometric meaning in the formal limit
       $\hbar \, \rightarrow \, 0$  and  $N \, \rightarrow \, \infty$.    For example, 
   in Einstein theory, in such a limit $r_*$ becomes a geometric Schwarzschild radius. 
   However, we shall not use this meaning.  Treating  $r_*$ quantum-mechanically 
   shall allows us to bypass unnecessary details emerging in exact classical limit, which 
   in any case represent an idealized approximation of the real world and are only  approximately  applicable to a realistic scattering processes. 
   
     Following the large-$N$ quantum portrait of black holes and classicalons,  for us $r_*$ 
     will be the scale (the characteristic wave-length)  for which the interaction of $N$-gravitons becomes strong and the system enters the regime in which a self-sustained bound-state starts to form.   
     
   %   However, we shall never need to go into the subtleties of this regime.  
    This picture   allows us to draw important conclusions  about 
    the quantum dynamics of this self-sustained bound-state without going beyond the  linearized approximation on the IR side.     
%      For Einstein gravity 
    %  the $N$-portrait has been intrioduced in\cite{giacesar}. The purpose of the  
 %  present work is to explore a wider landscape of  spin-2 classicalizing theories and   
    % generalize the $N$-portrait for a simple phenomenological reason to show that these represent 
   %  interesting probes for cosmic ray physics.  
      
       The key is to identify the energy dependence of the cross section.  As we shall show this is determined by the number and the couplings  of gravitational species 
       that $\delta g_{\mu\nu}$ propagates.   The number of new graviton species $n$ which 
 is a fixed  input characteristics  of a given theory,   should not be confused  with the occupation number,  
 $N$, of graviton quanta in a given black hole. The latter number depending on the black hole mass and takes different values for different black holes  in one and the same theory.      
       
          The most general expansion of the {\it linearized} metric perturbation in terms of graviton species has  the following form,   
   \begin{equation}
   \delta g_{\mu\nu} \, = \,  {1 \over M_P} \,  \sum_m  \, g^{(2)}_{m} \,  h_{\mu\nu}^{(m)} \,  \, + \, 
    {1 \over M_P} \,  \sum_m  \, g_{m}^{(0)} \,  \eta _{\mu\nu} \, \phi^{(m)}  \, , 
   \label{modeexp}
   \end{equation}   
  where $h_{\mu\nu}^{(m)}$  and   $\phi^{(m)}$ stand for spin-2 and spin-0 degrees of freedom respectively.  The constants $ g^{(2)}_{m}$ and  $g^{(0)}_{m}$ parameterize the strength of their coupling relative to the zero-mode Einstein graviton ($h_{\mu\nu}^{(0)}$), for which we have $g_0^{(2)}\, = \, 1$ \footnote{ 
  Notice, that since we couple the metric perturbation only to the conserved energy momentum tensor, 
  the would-be contribution from the derivatively-coupled scalars of the form 
   $\partial_{\mu}\partial_{\nu} \phi^{(m)}$ as well as the contributions from spin-1 states, vanishes.
    These are therefore neglected in the expansion (\ref{modeexp}).}.

 Consider now a localized static source of some characteristic  wave-length $\lambda = R$. 
Classically, such a source can be approximated by a static energy momentum tensor smeared 
over a spherical shell of radius $R$, 
  $T_{\mu\nu} \, = \, \delta_{\mu}^0\delta_{\nu}^0 {E \over 4\pi R^2} \delta(r-R)$.   For such a source,
  the linear graviton perturbations can be easily obtained by solving appropriate linearized equations which are well-known and we shall not display. 
 In particular,  for the Newtonian components we  get,      
\begin{equation}
\label{hfromM}
h_{00}^{(0)}  \, = \,    { L_PE \over 2 r} \,,  ~~~
h_{00}^{(m)}  \, = \, {2 \over 3}  g_{m}^{(2)} \,  { L_PE \over r} e^{-mr}  \, ,
~~~\phi^{(m)}  \, = \,   g_{m}^{(0)} \, { L_PE \over r} e^{-mr} \, .  
\end{equation}
 Substituting this expression into  (\ref{modeexp}) we get for the total metric perturbation the following expression, 
   \begin{equation}
   \delta g_{00} \, = \,  {L_P^2 E  \over r}   \sum_m  \, \rho_{m} \, e^{-mr} \, ,  
   \label{geff}
   \end{equation}   
  where
  % $r_g \, = \, 2 G_NM$ is the usual Einsteinian gravitational radius, and 
  $\rho_{m}$ parameterize the relative strengths of massive spin-$2$ states with respect to 
  the massless graviton.  That is,  $\rho_m \, \equiv \, {2 \over 3}  (g_m^{(2)})^2 \, + \, (g_m^{(0)})^2$, whereas $\rho_0 \, = \, 1/2$.  
   
   It is clear that the sum  in (\ref{geff}) measures the relative strength of gravity with respect to pure Einstein gravity, and can be written as, 
   \begin{equation}
   {\delta g_{00} \over \delta g_{00}^{(Einstein)}  } \, = \,  \sum_m  \, \rho_{m} \, e^{-mr} \, .  
   \label{factor}
   \end{equation}   
   
   The  quantum-mechanical picture described by the metric perturbation (\ref{geff}) is that 
  a source of energy $E$ induces a state of graviton occupation number,  
 \begin{equation}
   N \, = \,  \sum_m   N_m   \, = \,   (E^2 \, L_P^2)  \sum_m   \rho_m  \, e^{-mR}  \, .
  \label{Ngeneral}
  \end{equation}
   Thus, each graviton of mass $m$ and coupling $\rho_m$ contributes the  partial number 
 $N_m \,  = \, N_0 \,  \rho_m  e^{-mR}$, where $N_0 \, \equiv \,  E^2 L_P^2$ is the occupation number of the massless 
 gravitons. 
 
    In our parameterization the classicalization self-completion threshold is set by the  source of the Compton wave-length $L_*$ that creates an order-one  metric perturbation  at the distance $ r \sim L_*$. 
    That is,   $\delta g_{00}(r=L_*)  \, =  \ 1$.   Or equivalently,  in the quantum language 
 this is a source for which the occupation number of gravitons, $N$,  exceeds  one.   In other words, such 
 a source marks the boundary between the one-particle and multi-particle states.  
    
   Let us see, what relation this imposes on the spectrum of gravitational messenger species. 
   Since each graviton species of mass $m$ contributes only at distances 
   $r \, < \, 1/m$,  at any given distance $r$ we  are allowed to only take into the account 
   modes that are not heavier than $1/r$.   Thus, for the estimate we can truncate the sum at $m = 1/r$. 
    At the same time we can approximate the exponential factors for all the light modes 
    by one.  The sum then simplifies to, 
      \begin{equation}
   \delta g_{00} (r)  \, = \,  {r_g  \over r}   \sum_{m=0}^{m = 1/r}   \, \rho_{m}  \, .  
   \label{gsimple}
   \end{equation}   
  Now, the criterion that sources heavier than  $L_*^{-1}$ must classicalize  reads 
       \begin{equation}
   \delta g_{00} (r = L_*)  \, = \,  {L_P^2   \over  L_*^2}   \sum_{m=0}^{m = 1/L_*}   \, \rho_{m}  \, = \, 1 .  
   \label{central}
   \end{equation}   
   The bottom-line is that the cumulative strength of all the new degrees of freedom 
  at the scale $L_*$  must be $L_*^2/L_P^2$ times the strength of Einstein's gravity. 
   This condition only constraints the overall sum of  coupling strengths as given by (\ref{central}). 
   This constraint can be accommodated  in different ways.  For example, one can introduce 
  many weakly-coupled gravitons  or  few strongly coupled ones.   By changing 
  coupling strengths $\rho_m$ and the number of gravitons we can scan an entire 
  landscape of classicalizing  spin-2 theories.  
  
   Sources of energy $E \, \gg \, M_*$ classicalize and represent $N$-particle states. 
 This applies both to static sources  as well as  to scattering  two-particle states  with center of mass 
 energy $E$. 
  Our task is to single out the class of theories for which the classicalization radius can exhibit 
  the most efficient growth  at  phenomenologically-interesting low energies i.e above 
$M_* \sim $TeV.   This task is simple to accomplish. Each member of the graviton tower of mass 
$m_j$ and of strength $\rho_{m_j}$ contributes into 
the classicalization  process at  distances shorter than its Compton wavelength $m_j^{-1}$ 
and decouples at larger distances.   Since the integrated over-all strength of gravitons is fixed by 
(\ref{central})  the most efficient growth of $r_*(E)$ with $E$ will be achieved if we make all the new 
gravitons maximally light.  

   The classicalization radius of  any source of energy $E\, \gg \, M_*$ is determined by the relation,    
   \begin{equation}
 \delta g_{00} (r_*) \, = \,   {E\over M_P^2 r_*}  \left [ \sum_j \, e^{-m_j r_*} \,  \rho_{m_j} \, + \, 1 \right ] \, \simeq \, 1\, 
\label{Rg}
\end{equation}
 We now wish to find such consistent values of $m_j$ and $\rho_j$ that would ensure the 
 largest possible value of $r_*$ for a given mass $E$.  This is obviously achieved in the case 
 when most of the graviton masses are pushed down as much as possible. 
  At any given energy $E$ the gravitons contributing efficiently into the growth of $r_*$
  will be the ones that satisfy,  
\begin{equation}
 m_j \, <  \, m \, = \,  {M_*^2 \over E}  \, . 
\label{masses}
\end{equation}
 The extreme case is achieved when all the gravitons open up at the same universal mass $m$.    
 Since what is important is the integrated strength, we can take all the graviton species  to be  coupled with the uniform  strength set by their inverse number, $n$,  
\begin{equation}
 \rho_m  \, = \, L_*^2 /L_P^2 {1 \over n} \, .
\label{couplings}
\end{equation}
%where we have defined a scale $L_{BH}$.
 In other words,  for a given center of mass energy $E$, the largest classicalization 
 radius $r_*(E)$ is achieved  when all the  $n$ graviton species open up 
at the common Compton wavelength $m^{-1} \, \gg \, L_*$. 
 %The picture is then the following.   For  $E \, < \, M_*$, the scattering is perturbative. 
 % Starting from $E \, > \, M_*$  classicalization sets in, and sources of  energy 
 % $E$,  
 %such that  the latter exceeds or is of order of the scale $ E/M_*^2$. 
The  $r_*(E)$  radius of such a classicalon is then determined by  the condition: 
\begin{equation}
  {E\over M_*^2} {1 \over r_*} \,  \left [ e^{-m r_*} \, + \, {M_*^2 \over M_P^2} \right ] \, = \, 1
\label{Rg}
\end{equation}

In order to translate this result into the relation  between the classicalon  production cross-section ($\sigma$) and the center of mass energy ($E$), all we need to do is to identify  $\sigma \, = \, \pi r_*^2$ and plug it 
in (\ref{Rg}).   This identification gives us the relation, 
\begin{equation}
E \, = \, M_*^2  \, {\sqrt{{\sigma \over \pi}} \over  e^{-\sqrt{{\sigma \over \pi}}m} \, + \, M_*^2/M_P^2} \, .
\label{Esigma}
\end{equation}
 
 We can further simplify the above relation by noticing that for the values of interest $M_* \sim$TeV,  the term  $M_*^2/M_P^2$ is of order $ 10^{-32}$  and is absolutely negligible.  
The value of  $r_*(E)$ is then determined  by 
the condition
\begin{equation}
 {E\over M_*^2 }   \, { e^{-m r_*} \over r_*} \, = 1 
\label{Rg1}
\end{equation}
or equivalently, 
\begin{equation}
E \, = \, M_*^2  \, {\sqrt{{\sigma \over \pi}} \over  e^{-\sqrt{{\sigma\over \pi}}\, m}} \,.
\label{Einm}
\end{equation}
Which in terms of cross-section gives us (\ref{sigma}). 
For $m \, <  \, (EL_*^2)^{-1} $ the value of $r_*$  is given by 
\begin{equation}
r_* \, \simeq \,  E \, L_*^2  \,  \left (1 \, - \,  m E L_*^2 \right) \, ,
\end{equation}   
which in terms of cross-section gives us leading expansion of (\ref{sigma}) in $m E L_*^2$.  
%\begin{equation}
%\sigma  (E)  \, \simeq \,  \pi \, {E^2 L_*^4}  \, ln^2 \left ({1 \over  1\, - \, mEL_*^2}\right)
%\end{equation}   
  The latter expression is no longer applicable after the center of mass energy reaches 
  $E_c\, = \, (mL_*^2)^{-1}$, or equivalently the  $r_*$ radius reaches $m^{-1}$. Above this energy 
  we have to use the expression (\ref{Einm}) (which implies (\ref{sigma})), which indicates that the growth of cross section 
  becomes logarithmic,  
\begin{equation}
  \sigma \, = \, \pi m^{-2}  ln^2 \, (mEL_*^2)\, .
  \label{freeze}
  \end{equation}
 As it is clear from (\ref{Esigma}),  this  logarithmic growth persists till the energies of order $ E \, \sim \, M_P^2/m$, after which one enters into the regime in which  classicalization is taken over by  the Einsteinian massless graviton via production 
 of Einsteinian black holes.   Obviously,  this regime is way beyond the reach of current cosmic ray experiments. 
So the signature that is of interest for the current observations  is given by the cross section 
(\ref{sigma}).  As we shall show  in the next section, the phenomenological lower bound  on 
$m$ is as low as $\sim \, $MeV, allowing the classicalization cross section  to compete with or even dominate the QCD cross section at the energies around $\sim \, 100$TeV.

\section{Phenomenological Constraints on $m$}

 As we have shown above,  the 
 efficiency of the cross section growth in deep-UV is determined by the 
Compton wave-length of the gravitons.  This is a very peculiar characteristics of classicalizing theories and is not present in weakly-coupled Wilsonian UV completions.
 For example, the behavior of  the cross section at energy $E \, \gg \, L_*^{-1}$ instead  of  being  determined by the strength  of the interaction at  distance $\hbar/E$, is  rather  determined  by the strength 
 at distance $r_* \, \gg \, L_*$.  The gravitons of Compton wavelength shorter than $r_*$ effectively do not  participate in formation of classicalons. 
   Thus  for  the efficient growth of the cross section at energy $E$ it is important to have as many light  gravitons as possible with the Compton wavelength exceeding $r_*(E)$. 
   However,  graviton masses and couplings are constrained from below by various observations. 
  In case of large extra dimensions this constraints have been studied in great detail \cite{ADD1}. 
   We shall now apply the similar analysis in our more general case. 
    
  The number and couplings of gravitons available at each mass level $m$ are constrained
  by various high-energy processes as well as by the tabletop measurements. 
  We shall now briefly evaluate these constraints.  
  
  \subsection{Star Cooling} 
  
     We shall consider the star-cooling constraints first.  These constraints come from  the fact that
     gravitons can be produced in the interiors of the stars and contribute to their cooling.  
      The requirement that this is not disturbing the usual cooling rate,  places the 
     constraint on the masses and couplings of the graviton species.  
     
      We can consider two different regimes depending 
     whether mean free path of produced gravitons is  longer or shorter  than the star radius $R_{star}$.   In the first case the gravitons  escape freely, whereas in the second case  they can deposit 
     the energy back into the star before escaping.  
   We shall consider a free-escape regime first. 
   
   \subsubsection{ Free escape regime}

   In such a regime the star-cooling rate is essentially given by the graviton production rate, since probability of the re-capture is small.

     For a star with an interior temperature 
$T$,  the cooling rate due to  emission of massive gravitons is, 
\begin{equation}
\Gamma_{star \, \rightarrow \, graviton} \,  \sim \, {T^3 \over M_P^2} \, \sum_j \, \rho_j \,  e^{-m_j/T} \, ,
\label{starrate}
\end{equation}
(where for simplicity we do not make distinction between the spin-2 and spin-0 cases). 
The exponent comes from the Boltzmann suppression. 
This rate must be less than the standard cooling rate.  
The most stringent bound comes from  supernova 1987A, with the core temperature  $T \, =\, 30$MeV.  In order to make a quick estimate, we can use the knowledge from  the 
previous analysis  (such as production of axion \cite{axion} 
or   Kaluza-Klein gravitons \cite{ADD1}) indicating that the  cooling rate into new states should not exceed 
the quantity 
\begin{equation}
\Gamma_{max} \,  \sim \, {T^3 \over  10^{18} GeV^2}  \, . 
\label{maximal}
\end{equation}
Demanding that (\ref{starrate}) does not exceed (\ref{maximal}) gives  the following constraint, 
\begin{equation}
\sum_j \, \rho_j \,  e^{-m_j/30 MeV} \,  < \,  10^{20} \, .
\label{starconstraint}
\end{equation}
Let us recall,  that according to  (\ref{factor}) the sum entering on the right hand side of (\ref{starconstraint})
measures the strength of gravity relative to pure Einstein at distances  $r \,  = \,  (30MeV)^{-1}$.   
This means that  the star-cooling constraints could still tolerate modified gravity that at distances 
$r \, \sim  \, (30 MeV)^{-1} $ could become $10^{20}$-times stronger  relative to standard 
Einstein gravity.   With this constraint in mind let us evaluate the case which could give the largest 
possible cross section in the range of energies that  can be potentially probed by high energy cosmic ray experiments. 

 The two choices of the parameters that tend to enhance the classicalon production  cross section are:  1)  Pushing the scale  $M_*$ as low as possible; and 2) increasing the Compton wavelengths of gravitons as much as possible. So the maximal effect is achieved when almost all  the new graviton species open up 
 at the same mass scale $m$. 
  
   In such a case,  the ratio of the supernova cooling rate into the gravitons to the maximal 
 allowed cooling rate (\ref{maximal}) becomes, 
\begin{equation}
{\Gamma_{star \, \rightarrow \, graviton} \over \Gamma_{standard} } \,  \sim \, \left ({10^{9} GeV\over M_*}
\right )^2 \, e^{-m/30 MeV} \, .
\label{starratenew}
\end{equation}
LHC experiments already set the lowest bound on scale $M_*$ in the $\sim $ TeV range.  
 Substituting this lower limit in (\ref{starratenew}) and demanding that 
 the ratio must be below one  we arrive to the lower bound on  
 $m$ of few hundred MeV. 

\subsubsection{Energy recapture regime}

We shall now consider  a different regime  in which the individual gravitons are strongly enough coupled so that their mean free path is less than the star radius.  
  Such a situation, for example, can take place when gravitons that are 
 produced in a hot inner core of the star decay into the ordinary particles 
 before reaching the surface.  
 In such a case they deposit energy in the interior of the star before escaping and this suppresses  their contribution into the cooling.   
  Let us evaluate constraints for such a regime.   The lifetime  of a graviton of mass $m_j$ and 
  coupling $\rho_j /M_P^2$  produced at temperature  $T \, \gtrsim \, m$ in the interior of the star
 is 
 \begin{equation}
 \tau \, \sim \, {T \over m_j} \, {M_P^2 \over \rho_j \, m_j^3} \, .
 \label{lifetime}
 \end{equation}
 The coefficient $T/m_j$ comes from a relativistic gamma factor.   This lifetime sets the distance 
 that gravitons travel before decaying. Demanding that this distance is less than 
 the radius of a proto-neutron star,   $\tau \,  < \, R_{star} \, \sim \, 10$km, we get the following bound 
 on the mass, 
  \begin{equation}
 m_j^4  \, \gtrsim  \, {T M_P^2 \over R_{star} \rho_j } \, . 
 \label{mbound}
 \end{equation}
  For the case, in which all the graviton species have the same mass $m$, we can translate the 
  above bound in terms of their  number, $n$,  using the relation  (\ref{couplings}), 
  \begin{equation}
 m^4  \, \gtrsim  \, {T M_*^2n \over R_{star}  } \, . 
 \label{mn}
 \end{equation}
 This has a clear physical meaning.  For  less number of graviton species, in order  to accommodate the same collective strength,   the individual couplings must be stronger.  Thus,  the probability of  re-capture for each graviton  becomes  higher. 
 As an extreme case we can take $n=1$.  In this case taking $M_* \sim $TeV, we get the 
 following re-capture bound,  
 \begin{equation}
  m \, \gtrsim \, MeV \, .
 \label{onebound}
  \end{equation}
   Notice that in  approximating the  mean free path by (\ref{lifetime})  we assumed  
 that the particles into which  gravitons decay are off-shell (due to thermal bath)  by a sufficiently small amount  so that the decay is still allowed.  This is a reasonable approximation for the current estimate. 
 
  To summarize,  the supernova gives  the following constraints on the scale $m$  in two different regimes.    In a free-streaming regime, when gravitons are many and weakly coupled, the
 lower  bound is about of few $100$MeV.   In the recapture regime, when gravitons are less and strongly coupled, the bound is $\sim MeV$.  

\subsection{Other Atro-Cosmo Constraints} 

 With the above bounds being satisfied the other astro-cosmo constraints (e.g., such as nucleosynthesis or the diffused gamma ray background) are either automatically satisfied or are much milder.  
 
    Let us consider for example a big bang nucleosynthesis (BBN)  bound. 
  This bound can be estimated from the requirement that graviton production rate 
  at the BBN  temperature $T_{BBN} \, \sim \, $MeV,  should be subdominant to the expansion rate of the   Universe which in that epoch is  $H_{BBN} \, \sim  \,  T_{BBN}^2/M_P$. This requirement 
  gives the bound, 
\begin{equation}
 \sum_j \, \rho_{m_j} \,  e^{-m_j/MeV} \,  < \,  10^{22} \, .
\label{BBN}
\end{equation}
This bound is milder than (\ref{starconstraint}) and thus is automatically 
satisfied when supernova cooling  constraint is met in the free-streaming regime.   For example, when all the graviton species  have a common mass $\sim  100$ MeV.  

  In the recapture case, when the number of graviton species  is smaller and they are stronger coupled,   BBN bound is 
  slightly stronger than the supernova bound and gives $m \, \gtrsim \, 20$MeV.

  Given the constraints from supernova and BBN the constraints from the diffused photon background are easy  to accomodate.  This constraint usually comes from a possible late decay  of the gravitons\cite{ADD}. 
  
   Even if produced with a maximal abundance,  for  the value  of $m$ even at its lowest bound
 $20$ MeV, the lifetime of gravitons can be easily made shorter  than the age of the Universe.   Indeed, for number of graviton species being $n$, the  lifetime of graviton of mass 
 $m$ is,  
 \begin{equation}
 \tau_{gravitons} \, \sim  \, M_*^2n/m^3 \, .
\end{equation}
Requiring that this is less than the age of the Universe, $\tau \, < \, H_0^{-1} \sim 10^{28}$cm, we  get a very mild upper constraint on the total number of gravitational species $n \,  \lesssim \, 10^{32} $,  which in any case 
is satisfied due to the constraints on the number of gravitational species implicit in equation (\ref{central}).

\subsection{Tabletop Measurements} 

 Other constraints on the Compton wavelengths of new graviton species come from 
 tabletop measurement of deviations from Newton's law. 
  These experiments have been probing distances down to the micron range
  and are placing bounds on gravity competing forces of various strengths. 
  
  In case when all the new gravitons are heavier than the star cooling constraint, the Compton wavelengths are too short to be of interest for the above-mentioned table top measurements. 
  
   However, as we have shown above, the combined star-cooling and BBN constraints allow existence of longer Compton wave-length gravitons as long as their cumulative strength
   is less than $\sim \, 10^{20}$ times the strength of Einstein gravity.  This bound leaves a huge room 
   for the existence of lighter gravitons that could be of interest for table top 
   experiments.   The absolute upper bound on Compton wave-length of any new graviton 
   interacting with approximately the same (or larger) strength as a massless graviton, 
   comes from the torsion balance experiment  and is  $0.5$ mm \cite{adelberger}. 
    Shorter scale experiments are sensitive to stronger forces.

\section{Concluding Remarks} 

 The point of this note was to outline the potential importance of  classicalizing theories for ultra-high energy experiments with relatively low luminosity, such as the high-energy cosmic ray 
 or plasma Wakefield accelerators\cite{Allen}.  
 We have identified some phenomenologically-consistent theories that exhibit 
  a most efficient  growth of the cross section at the energies of experimental interest. 
 
   We have stressed importance of underlying quantum picture \cite{number}  according to which the black holes (or classicalons in general) are Bose condensates of maximal (for a given size) occupation number $N$.
   All the characteristics of this system can be understood in terms of $N$. 
  For the phenomenological studies in the ultra-high cosmic ray energy range,  this quantum picture is important because of at least two reasons. First, it allows to map the linear analysis   
  on an underlying quantum portrait, which would be impossible in 
  the classical limit $N = \infty$, $r_* = fixed$. 
  Secondly, it allows to cross correlate expected events at LHC with the ones 
  in ultra-high energy searches.  For example, in the considered classicalizing models
  within the energy interval between  $\sim  1-100$ TeV  we scan the classicalons 
 states with occupation number ranging in an interval $N \, \sim \, 1 -10^4$.    
  Thus,  we predicts, that the two-jet events at LHC that are expected to come from the decay 
  of the lightest classicalon resonances ($N \, \sim \, 1$)
  must be cross-correlated with the multi-jet events in higher energy collisions.   
  
   We have identified a  phenomenologically-viable class of classicalizing spin-2 theories, 
 and have shown that  the current observational constraints put a limit  on the saturation point
 of the cross-section-growth at about $\sigma \, \sim \, (20 MeV)^2$.    This  cross 
 section dominates over the soft QCD cross section and thus represents an interesting potential 
 target for the experiments with high energy but low luminosity collisions, such  
 that are expected in cosmic rays or at  plasma wakefield accelerators.

\section*{Acknowledgements}

  It is a great pleasure to thank Allen Caldwell for sharing his expertise on Plasma Wakefield accelerators and their prospect for  particle physics.  
  We also gratefully  acknowledge earlier discussions with Masahiro Teshima and Alexey Boyarsky on topics related to physics of  micro black holes in cosmic rays.  
The work of G.D. was supported in part by Humboldt Foundation under Alexander von Humboldt Professorship,  by European Commission  under 
the ERC advanced grant 226371,  by European Commission  under 
the ERC advanced grant 226371,   by TRR 33 \textquotedblleft The Dark
Universe\textquotedblright\   and  by the NSF grant PHY-0758032. 
The work of C.G. was supported in part by Grants: FPA 2009-07908, CPAN (CSD2007-00042) and HEPHACOS S2009/ESP-1473 and by  Alexander von Humboldt Foundation.


\begin{thebibliography}{99}

 \bibitem{Rev}
 
A.W.~Strong, I.V.~ Moskalenko, and V.S.~ Ptuskin, "Cosmic-Ray Propagation and Interactions in the Galaxy", Annual Review of Nuclear and Particle Science
Vol. 57: 285-327 



  \bibitem{Allen}
    A.~Caldwell,  "Proton-driven plasma Wakefield Acceleration",  talk given at CERN colloquium (2012), 

http://indico.cern.ch/conferenceDisplay.py?confId=182516; 

For the discussions about the Plasma Wakefield Acceleration as giving the possibility of reaching high energies but with lower luminosities than discussed for ILC or CLIC, see, 

A. Caldwell, K. Lotov Phys. Plasmas 18, 103101 (2011)

T. Tajima, M. Kando, M. Teshima, Prog.Theor.Phys. 125 (2011) 617-631




 \bibitem{gia-cesar}


G.~Dvali and C.~Gomez, Self-Completeness of Einstein Gravity,  arXiv:1005.3497 [hep-th]; 

G. Dvali, S. Folkerts, C. Germani, ÒPhysics of Trans-Planckian GravityÓ, [ arXiv:1006.0984 [hep-th]]
Phys.Rev.D84:024039,2011. 




  \bibitem{class}

G.~ Dvali, G.~ F. Giudice, C.~ Gomez, A.~ Kehagias,  UV-Completion by Classicalization,
JHEP 2011 (2011) 108; arXiv:1010.1415 [hep-ph].  

For recent discussions, see




\bibitem{giadavid} G.~Dvali, D.~Pirtskhalava, 
Dynamics of Unitarization by Classicalization, 
  Phys.Lett. B699 (2011) 78-86,  arXiv:1011.0114 [hep-ph]

G.~Dvali,
 Classicalize or not to Classicalize?,   arXiv:1101.2661 [hep-th]

\bibitem{regularBH}	
E.~Spallucci, S.~ Ansoldi, 
Regular black holes in UV self-complete quantum gravity, 
Phys.Lett. B701 (2011) 471-474,  arXiv:1101.2760 [hep-th]

	

\bibitem{cosmclass1} F.~ Berkhahn, D. D. ~Dietrich, S.~Hofmann,  
Cosmological Classicalization: Maintaining Unitarity under Relevant Deformations of the Einstein-Hilbert Action,  Phys.Rev.Lett. 106 (2011) 191102,   arXiv:1102.0313 [hep-th]

F.~ Berkhahn, D. D.~ Dietrich, S.~ Hofmann 
Consistency of Relevant Cosmological Deformations on all Scales.
JCAP 1109 (2011) 024 
e-Print: arXiv:1104.2534 [hep-th]

F. ~Berkhahn, S.~ Hofmann, F.~ Kuhnel, P.~ Moyassari,  Dennis Dietrich,  
Island of Stability for Consistent Deformations of Einstein's Gravity.  arXiv:1106.3566 [hep-th]




\bibitem{borutgoran} 

B.~ Bajc, A.~ Momen, G.~ Senjanovic,  
Classicalization via Path Integral,  arXiv:1102.3679 [hep-ph]


\bibitem{gia-cesar-alex}

G.~Dvali, C.~Gomez, A.~Kehagias, Classicalization of Gravitons and Goldstones.
arXiv:1103.5963 [hep-th],  JHEP 1111 (2011) 070
	


\bibitem{phenoclass} 

C.~ Grojean,  R. S. Gupta, Theory and LHC Phenomenology of Classicalon Decays, 
 arXiv:1110.5317 [hep-ph].


	
\bibitem{tetradis}
J. Rizos, N. Tetradis,  Dynamical classicalization, 
 arXiv:1112.5546 [hep-th]

	
N. Brouzakis, J. Rizos, N. Tetradis,  On the dynamics of classicalization, 
Phys.Lett. B708 (2012) 170-173,   arXiv:1109.6174 [hep-th]


\bibitem{noclass}
R.~ Akhoury, S.~ Mukohyama, R.~Saotome, 
No Classicalization Beyond Spherical Symmetry,  arXiv:1109.3820 [hep-th]

\bibitem{sigmamodel} 	
R.~ Percacci, L.~ Rachwal,  
On classicalization in nonlinear sigma models, 
 arXiv:1202.1101 [hep-th]


\bibitem{number}
G.~Dvali, C.~Gomez, Black Hole's Quantum N-Portrait, arXiv:1112.3359 [hep-th]; 	
Landau-Ginzburg Limit of Black Hole's Quantum Portrait: Self Similarity and Critical Exponent, 
arXiv:1203.3372 [hep-th]; 	
Black Hole's 1/N Hair, arXiv:1203.6575 [hep-th]. 

\bibitem{tHooft} Gerard 't Hooft,  A Planar Diagram Theory for Strong Interactions.
 Nucl. Phys. B72 (1974) 461


\bibitem{Witten}
E.~ Witten, 	
Baryons in the 1/n Expansion. Nucl.Phys. B160 (1979) 57

\bibitem{transplanck} G.~ Õt Hooft, Phys. Lett. B198, 61-63 (1987).




\bibitem{ADD}
  N.~Arkani-Hamed, S.~Dimopoulos and G.~R.~Dvali,
  ``The hierarchy problem and new dimensions at a millimeter,''
  Phys.\ Lett.\  B {\bf 429}, 263 (1998)
  [arXiv:hep-ph/9803315];

\bibitem{AADD}
 I.~Antoniadis, N.~Arkani-Hamed, S.~Dimopoulos and G.~R.~Dvali,
  ``New dimensions at a millimeter to a Fermi and superstrings at a TeV,''
  Phys.\ Lett.\  B {\bf 436}, 257 (1998)
  [arXiv:hep-ph/9804398].
 
 \bibitem{micro}

 G.~ Dvali, G. Gabadadze, M. Kolanovic and F. Nitti, Phys. Rev. D 65 (2002) 024031
[arXiv:hep-th/0106058]; 

S.B.~ Giddings and S.D.~Thomas, Phys. Rev. D 65 (2002) 056010
[arXiv:hep-ph/0106219]; 

S.~Dimopoulos and G.L.~Landsberg, Phys. Rev. Lett. 87 (2001) 161602
[arXiv:hep-ph/0106295]. 


 
 
  
\bibitem{tH}
  G.~'t Hooft,
 ``Graviton Dominance in Ultrahigh-Energy Scattering,''
  Phys.\ Lett.\  {\bf B198}, 61-63 (1987).

\bibitem{ACV}
  D.~Amati, M.~Ciafaloni, G.~Veneziano,
  ``Superstring Collisions at Planckian Energies,''
  Phys.\ Lett.\  {\bf B197}, 81 (1987);
``Classical and Quantum Gravity Effects from Planckian Energy Superstring Collisions,''
  Int.\ J.\ Mod.\ Phys.\  {\bf A3}, 1615-1661 (1988);

\bibitem{Gross}
D. J.~ Gross, P. F.~ Mende,  Nucl. Phys. B 303 (1988) 407;   

\bibitem{ADD1}
N.~ Arkani-Hamed, S.~ Dimopoulos, G.R. Dvali, 
Phenomenology, astrophysics and cosmology of theories with submillimeter dimensions and TeV scale quantum gravity,  Phys.Rev. D59 (1999) 086004, hep-ph/9807344. 



\bibitem{axion} G. G. Raffelt, Lect. Notes Phys. 741, 51 (2008).




   \bibitem{adelberger} 

S.~ Schlamminger, K.-Y. Choi, T.A. Wagner, J.H. Gundlach, E.G. Adelberger, 
 Test of the equivalence principle using a rotating torsion balance,  Phys.Rev.Lett. 100 (2008) 041101;  arXiv:0712.0607 [gr-qc]


   
\end{thebibliography}
\end{document}